\documentclass[aps,prb,twocolumn,amsmath,amssymb]{revtex4}
\usepackage{graphicx}
\usepackage{dcolumn}
\usepackage{bm}

\begin{document}\setlength{\unitlength}{1mm}

\title{Density functional theory for a  model quantum dot: \\
Beyond the local-density approximation}

\author{S. Schenk, P. Schwab, M. Dzierzawa, and U. Eckern}
\affiliation{Institut f\"ur Physik, Universit\"at Augsburg, 86135 Augsburg, Germany}

\date{\today}

\begin{abstract}
We study both static and transport properties of model quantum
dots, employing density functional theory as well as
(numerically) exact methods. For the lattice model under consideration the
accuracy of the local-density approximation generally is poor. For weak interaction,
however, accurate results are achieved within the optimized effective potential method,
while for intermediate interaction strengths a method combining
the exact diagonalization of small clusters with density functional theory
is very successful. Results obtained from the latter approach yield very good
agreement with density matrix renormalization group studies, where the full
Hamiltonian consisting of the dot and the attached leads has to be diagonalized.
Furthermore we address the question whether static density functional
theory is able to predict the exact linear conductance through the dot correctly---with,
in general,
negative answer.
\end{abstract}

\maketitle
\section{Introduction} 
Density functional theory (DFT) is an efficient tool for determining the electronic structure of solids.
While originally developed for continuum systems with
Coulomb interaction,\cite{hohenberg1964,kohn1965} DFT has also been applied to lattice models, such 
as the Hubbard model or models of spinless fermions.\cite{gunnarsson1986,schonhammer1987,schonhammer1995,lima2003}
These lattice models often allow for exact solutions---either
analytically or based on numerics---which hence can
serve as benchmarks for assessing the quality of approximations.

Very popular in solid-state applications is the
local-density approximation (LDA) where the exchange-correlation energy of
the inhomogeneous system under consideration is constructed via a local
approximation from the homogeneous electron system. \cite{schwin2008}
Recently a lattice version of LDA has been suggested for
one-dimensional systems, where the underlying homogeneous 
system can be solved using the Bethe ansatz.
For example, it has been demonstrated that the Bethe ansatz LDA
describes well the low-frequency, long-wavelength excitations of the interacting
one-dimensional system, i.e., of a Luttinger liquid.\cite{schenk2008,dzierzawa2009}

On the other hand LDA often fails in correlated systems and systematic improvements beyond the LDA are difficult.
In this article we focus on a model of spinless fermions describing
interacting electrons on a quantum dot. In a first step
we compare the equilibrium properties of the system, i.e., the number 
of particles on the dot as a function of the gate voltage obtained within 
different approximations for the exchange-correlation energy: the LDA
and the optimized effective potential (OEP) approach. Furthermore we
suggest a novel method, where the exchange-correlation energy is
obtained via the exact diagonalization of a small cluster that is
composed of the strongly interacting region and a few additional
sites.

In the second step we compute the linear conductance through the dot.
A general motivation for this study is recent progress in the field of molecular
electronics, where DFT-based calculations are a standard tool to
calculate electrical conductances, 
\cite{brandbyge2002,rocha2006,koentopp2008} however the conceptional
limitations of the approach are not very well understood yet. More specifically
we were motivated by the model studies in Refs.\
[\onlinecite{schmitteckert2008}]
and [\onlinecite{mera2010}]. 
Mera {\em et al.}
\cite{mera2010} stressed that static DFT reproduces the
conductance of an interacting system correctly if there exists a Friedel sum
rule that relates the conductance with the equilibrium density.
Schmitteckert and Evers \cite{schmitteckert2008} compared
conductances obtained from a density matrix renormalization group
calculation with those obtained within static DFT. Close to
resonances both conductances were in very good agreement, while
off-resonance there was a considerable discrepancy. 
This discrepancy is due to a exchange-correlation contribution to the
voltage difference between the two reservoirs, \cite{koentopp2008} $U_{xc}$. 
One of the questions we will address is whether $U_{xc}$ depends on the
distance between the interacting region and the reservoirs.

In the following section we will introduce the model under
investigation. In Sec.\ III, devoted to static density
functional theory, we discuss the approximations used to obtain
the exchange-correlation energy. Section IV is devoted to transport: 
We rederive the Meir-Wingreen formula for the conductance starting
from the dynamical density-density response function, and we apply the
formula to calculate the conductance for our model. The final section
contains a summary as well as our conclusions.
\section{The model}
\begin{figure}
\includegraphics[width=8.5cm]{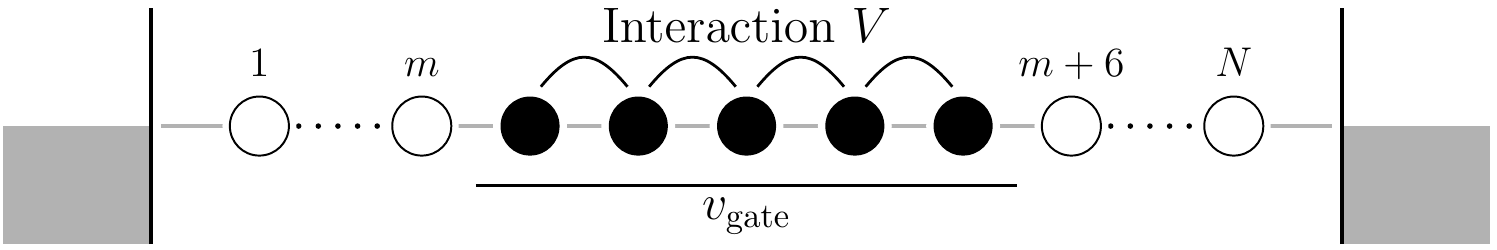} 
\caption{Schematic view of the model under consideration: A quantum dot (full circles) attached 
to left and right leads, each of them consisting of $m$ non-interacting sites (open circles) and a reservoir
described by a continuum of states (shaded regions).
The electrons on the dot (site $m+1$ to $m+5$) interact and their potential energy can be tuned by a gate voltage.
}
\label{fig-model}
\end{figure}

We study a model where a chain of $N$ lattice sites is coupled to two reservoirs
\begin{equation}
\hat H = \hat H_L + \hat H_{LC} + \hat H_C + \hat H_{CR} + \hat H_R 
.\end{equation}
The Hamiltonian of the chain reads
\begin{eqnarray}
\hat H_{C} & = & - \sum_{l=1}^{N-1} t_{l,l+1} \left(\hat c^+_l \hat c_{l+1} + \hat c^+_{l+1} \hat c_l\right) 
+ \sum_{l=1}^{N} v_l \hat n_l\nonumber \\
 & & + \sum_{l=1}^{N-1} V_{l,l+1} \left(\hat n_l - \frac{1}{2} \right) \left(\hat n_{l+1} - \frac{1}{2} \right)
\end{eqnarray}
where 
$\hat c^+_l$ and  $\hat c_{l}$ are fermion creation and annihilation
operators, $\hat n_l = \hat c^+_l\hat c_{l} $ counts the fermions on lattice site $l$.
The hopping matrix elements $t_{l,l+1}$ are chosen such that the
system resembles a quantum dot that is weakly coupled to left and right leads,
cf.~Fig.~\ref{fig-model}, and are explicitly given by
\begin{equation}
t_{l,l+1} = 
\left\{\begin{array}{ll}
                t                   & \  l = 1,\ldots,m-1, m+6, \ldots, N-1\\
                t'        & \  l = m \ \ \ {\rm and} \ \ \ l = m+5 \\
                t_{\rm dot}  & \  l = m+1,\ldots,m+4
       \end{array}
\right.
\end{equation}

In the following we choose $t' = 0.2t$ and $t_{\rm dot} = 0.5t$.
The interaction strength and the potentials are constant within the quantum dot,
$V_{l,l+1} = V$ and $v_l = v_{\rm gate}$, and zero outside.
The reservoirs, chosen to be non-interacting fermions, are described by
\begin{equation}
\hat H_{L/R} = \sum_k \epsilon_k \hat c_{k L/R}^+ \hat c_{k L/R}^{}
,\end{equation}
and finally the coupling between the chain and the left reservoir reads
\begin{equation}
\hat H_{LC} = - \sum_k t_k \left(\hat c_{kL}^+ \hat c_1^{} + \hat c_1^+ \hat c_{kL}^{}  \right)
\end{equation}
and analogously for $\hat H_{CR}$.
The hopping parameters $t_k$ are fine-tuned such
that an electron at the Fermi energy is not back-scattered at the
interface between the chain and the reservoirs, \cite{schmitteckert2008}
$t_k  = \sqrt{t/\pi N(\epsilon_k) } $ where $N({\epsilon_k }) $ is
the density of states in the reservoirs. 
Up to this point our model is the same as the model studied by
Schmitteckert and Evers 
[\onlinecite{schmitteckert2008}]. In contrast to
[\onlinecite{schmitteckert2008}], however, we replace the discrete
levels in the reservoirs by a
continuum of a wide and flat band, which then leads to
\begin{eqnarray} 
\Gamma_{L/R} &=  & 2 \pi  \sum_k | t_k |^2 \delta(\epsilon_F -
\epsilon_k)  \\
             & = & \left. 2 \pi |t_k|^2  N(\epsilon_k) \right |_{k=k_F}  =2t .
\end{eqnarray}
Notice that in the continuum limit the density of states in the reservoirs
goes to infinity, and thus---in order to keep
$\Gamma_{L/R}$ constant---the coupling strength between the states in
the reservoirs and the chain goes to zero.

\section{Static density functional theory}
\label{Sec:DFT}
The lattice version of DFT
relies on the fact that there is a one-to-one correspondence between
local potentials $\{ v_i \}$ and the ground-state expectation values
of the site occupations $\{ n_i \}$. Therefore it is---in principle---possible
to express all quantities that can be obtained from the
ground-state wave function as a function of the densities.
The site occupations as a function of the potentials can be found from
derivatives of the ground-state energy with respect to the local
potential
\begin{equation}
 n_i =  \frac{ \partial E_0 }{\partial v_i}.
\end{equation}
In order to determine the potentials from the densities it is
convenient to define the function
\begin{equation}
F(\{ n _i \} ) = {\rm min }_{\Psi \to \{ n_i \}  } \langle \Psi | \hat T + \hat V | \Psi \rangle,
\end{equation}
where $\Psi \to \{ n_i \} $ indicates that the minimization is
constrained to such wave functions $\Psi $ that yield the given site
occupations $\{ n_i \}$. Here $\hat T$ and $ \hat V$ are the kinetic and the
interaction parts of the Hamiltonian, respectively. The ground-state
energy is obtained by minimizing the function 
\begin{equation} 
\label{eofn}
E(\{ n_i \} ) = F(\{ n_i \} ) + \sum_i v_i n_i 
\end{equation}
with respect to
$n_i$. When we minimize $E$ under the constraint of a constant
particle number we obtain the potential up to an additive constant
(the Lagrange multiplier):
\begin{equation}
v_i = - \frac{\partial F}{\partial n_i} + \lambda
.\end{equation}

A major step towards the practical implementation of DFT is to employ a
non-interacting auxiliary Hamiltonian $\hat H^s$ (Kohn-Sham
Hamiltonian) in order to calculate the density profile,
\begin{equation}
\hat H^s = \hat T + \sum_i v_i^s \hat n_i^{} 
,\end{equation}
where the potentials $v_i^s$ have to be chosen such that in the
ground-state of $H^s$ the site occupations $n_i$ are the same as in the
interacting model.
In analogy to the interacting system, the ground-state energy of the
Kohn-Sham system is found by minimizing
\begin{equation}
\label{esofn}
E^s(\{ n_i \} ) = F^s(\{ n_i \} ) + \sum_i v_i^s n_i^{}.
\end{equation}
Combining (\ref{eofn}) and (\ref{esofn}) yields 
\begin{equation}
E(\{ n_i \} ) = E^s(\{ n_i \} ) + E^{\rm HXC}( \{ n_i \}) + \sum_i (v_i -
v_i^s) n_i,
\end{equation}
with the Hartree-exchange-correlation energy defined by
\begin{equation}
E^{\rm HXC} ( \{ n_i \} ) = F(\{ n_i \} ) - F^s(\{ n_i \}) 
.\end{equation}
The condition that both $E$ and $E^s$ are minimal for the same set of site occupations $n_i$ 
requires that
\begin{equation}
\label{kseq}
v_i^s = v_i + \frac{\partial E^{\rm HXC}}{\partial n_i}.
\end{equation}

Up to this point no approximations have been employed. However to
determine $E^{\rm HXC}$ at a given density exactly is as demanding as finding
the ground state energy for a given potential. The hope is that there
exist good approximations for $E^{\rm HXC}$ that are accessible with low
numerical cost but still allow good estimates for the ground-state
energy and density. Here and in the following we will compare three
different approximations: the local-density approximation, the optimized effective
potential (so-called exact exchange) approximation, and finally a method based on
the exact diagonalization of small clusters. 

\subsection{Local density approximation}
In the LDA one writes $E^{\rm HXC}$ as the sum of
the (non-local) Hartree energy plus an exchange-correlation energy 
which depends only on the local density,
\begin{equation}
E^{\rm HXC}_{\rm LDA}(\{ n_i \} ) =  
V \sum_i  n_i n_{i+1 } + \sum_i \epsilon_{\rm XC}(n_i)
.\end{equation}
The local exchange-correlation energy 
$\epsilon_{\rm XC}(n)$ is determined from the ground-state energy density of a
homogeneous system at the same density. For the one-dimensional
lattice models this quantity can be calculated using the Bethe ansatz,
see Refs.\ [\onlinecite{lima2003,schenk2008}].
Notice that in the Hamiltonian we study the interaction strength
depends on position, and an ambiguity arises how to determine the
exchange-correlation potential for those sites which
interact only with one neighbor.
For simplicity we used in our numerical implementation of the LDA the same function $\epsilon^{\rm xc}(n)$ for all
interacting lattice sites, i.e., number $m+1$ to $m+5$.

\subsection{Optimized effective potential}
In the OEP approach the Hartree-exchange-correlation energy is
\begin{equation}
\label{ehxcoep}
E^{\rm HXC}_{\rm OEP}(\{ n_i \} ) = V \sum_i  n_i n_{i+1} + E^{\rm X}(\{n_i \})
\end{equation}
with the Fock-like exchange energy
\begin{equation}
E^{\rm X} = -V \sum_i 
\langle \hat c^+_{i} \hat c^{}_{i+1} \rangle \langle \hat c^+_{i+1} \hat c^{}_{i} \rangle.
\end{equation}
We calculate the ground state expectation values $\langle \hat c^+_{i} \hat c^{}_{i+1} \rangle$ etc.~for a non-interacting system coupled to reservoirs using the Green's function technique, see section IV.
The Kohn-Sham equations (\ref{kseq}) are most conveniently solved by iteration.
Starting with an initial guess for the potentials $v_i^s$ we
calculate the corresponding site occupations $n_i$ and the Hartree-exchange-correlation energy  $E^{\rm HXC}$.
Since in the OEP approach $E^{\rm HXC}$ depends only implicitly on $n_i$ we rewrite Eq.~(\ref{kseq}) using
\begin{equation}
\frac{\partial E^{\rm HXC}}{\partial n_i} = \sum_j \frac{\partial E^{\rm HXC}}{\partial v_j^s} 
\frac{\partial v_j^s}{\partial n_i}
\end{equation}
where the derivatives of $E^{\rm HXC}$ with respect to the $v_j^s$ are calculated numerically 
and $\partial v_j^s/\partial n_i$ is obtained by matrix inversion from $\partial n_i/\partial v_j^s$.
Finally we obtain a new set of Kohn-Sham potentials $v_i^s$. The whole procedure is repeated until convergence,
i.e., until the difference between old and new potentials is smaller than some given cutoff.

The fact that we use a continuum of states to describe the reservoirs simplifies
the task to solve Eq.~(\ref{kseq}) considerably: Since the states in
the reservoirs are only infinitesimally weakly coupled to the dot,
the Hartree-exchange-correlation potential in the reservoirs disappears, so that the number
of potentials $v_i^s$ to be determined self-consistently equals the chain length $N$.

\subsection{Exact diagonalization}
In our model Hamiltonian electrons interact only in a spatially
confined region. In the non-interacting regions we find numerically
(e.g., within the OEP approach, see Fig.~\ref{voep} below; compare also
Ref.~[\onlinecite{schmitteckert2008}] for density matrix renormalization group
(DMRG) results) only small exchange-correlation
potentials.
This finding motivated us to use the exchange-correlation energy of a
small cluster consisting of the interacting region plus a small
number of non-interacting sites as an approximation for the
exchange-correlation energy of the system attached to reservoirs:
\begin{equation}
E^{\rm HXC}_{\rm ED} (\{ n_i \} ) = F_{\rm ED}^{}(\{ n_i \}) - F_ {\rm ED}^s( \{ n_i \}) 
\end{equation}
where $F_{\rm ED}^{}(\{ n_i \})$ and $F_{\rm ED}^s(\{ n_i \})$ are exact on the small cluster and can be obtained by numerical
diagonalization.

In this approach we have to fine-tune the local potentials of three
different Hamiltonians such that all the three yield the same local
densities:
(i) a cluster of interacting electrons with potentials $u_{i}$,
(ii) a cluster of non-interacting electrons with potentials $u_{i}^s$, and
(iii) the Kohn-Sham Hamiltonian of the extended quantum dot attached
to reservoirs with $v_i^s = v_i + v_i^{\rm HXC}$
where $v_i^{\rm HXC} =  u_{i} - u_{i}^s $ on the cluster sites and zero outside.
Again, in a practical scheme the fine-tuning procedure is performed by iteration.
The task to determine the potentials $u_{i}$ that correspond to a given set of site occupations $n_i$
is nontrivial for an interacting system and limits the cluster size in our approach to
approximately 12 to 14 sites.

\begin{figure}
\includegraphics[width=8.5cm]{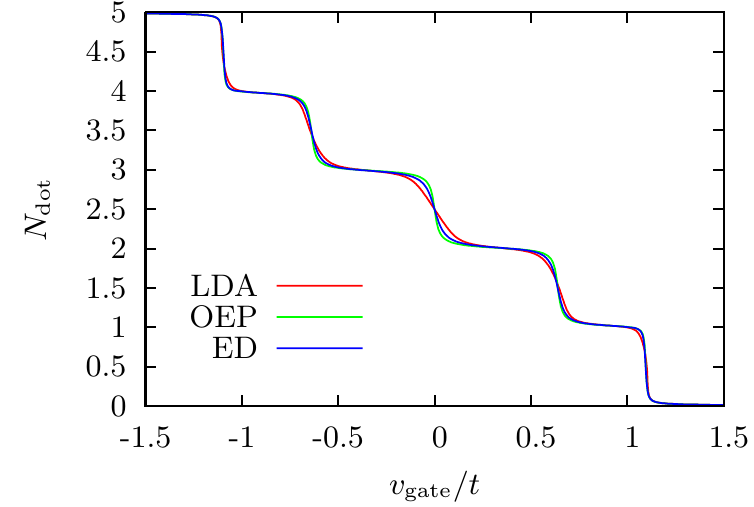} 
\caption{Particle number on the dot, $N_{\rm dot}$, as function
of the gate voltage, $v_{\rm gate}$, for $V/t = 0.25$ obtained within
density functional theory with three different approximations for the
exchange-correlation energy: the local-density approximation (LDA),
optimized effective potential (OEP) and a method based on the exact
diagonalization of short chains (ED). Here the chains have a length 
of nine lattice sites.}
\label{ndotV025}
\end{figure}
\begin{figure}
\includegraphics[width=8.5cm]{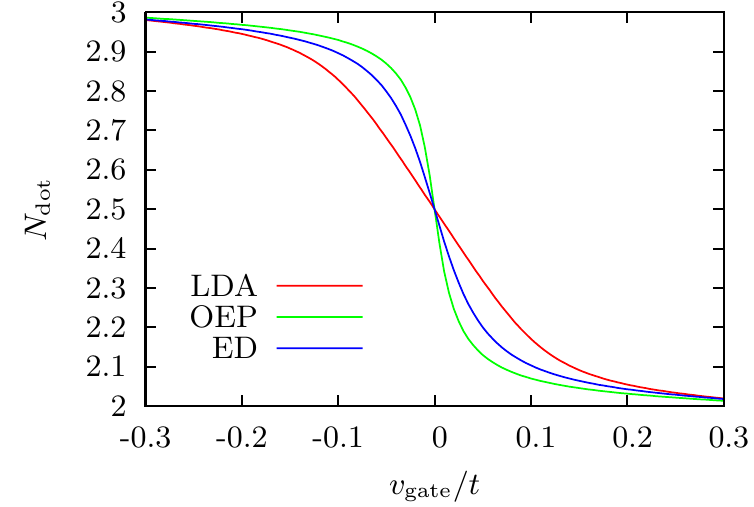} 
\caption{$N_{\rm dot}$ as function of $v_{\rm gate}$ around to the central step for $V/t = 0.25$; for this interaction strength
OEP is still close to the exact result.}
\label{ndotV025a}
\end{figure}
\begin{figure}
\includegraphics[width=8.5cm]{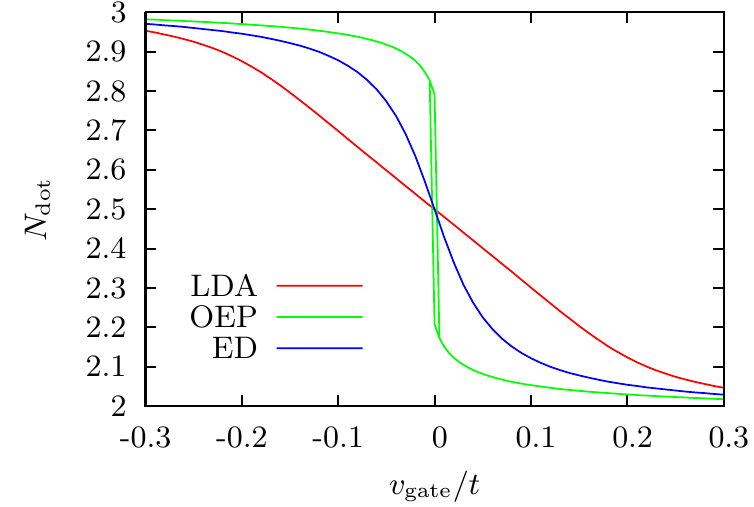} 
\caption{$N_{\rm dot}$ as function of $v_{\rm gate}$ for $V/t = 0.5$. 
Solving the OEP equations iteratively as a function of gate voltage two solutions are found close
to $v_{\rm gate} = 0$.}
\label{ndotV05}
\end{figure}
\begin{figure}
\includegraphics[width=8.5cm]{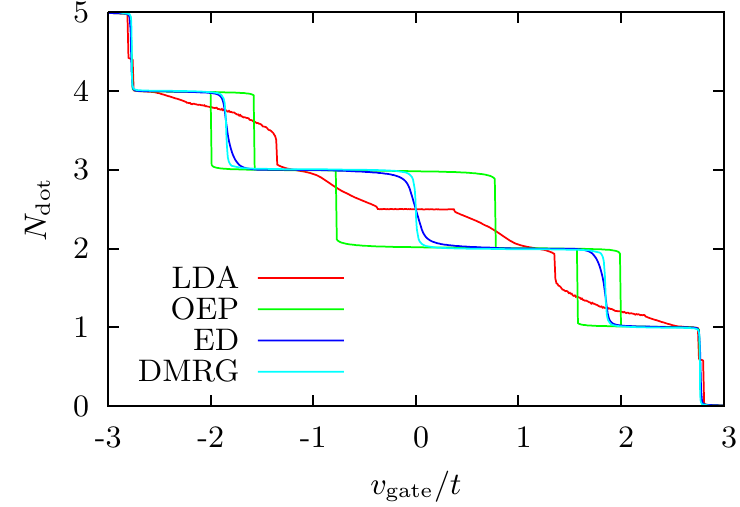} 
\caption{$N_{\rm dot}$ as function of $v_{\rm gate}$ for $V/t = 2$. 
For comparison we also include the DMRG results of [\onlinecite{schmitteckert2008}].}
\label{ndotV2}
\end{figure}
\begin{figure}
\includegraphics[width=8.5cm]{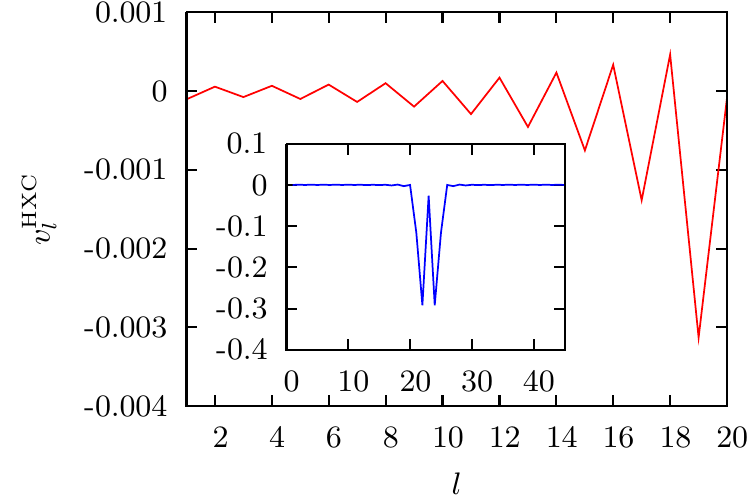}
\caption{Hartree-exchange-correlation potential $v^{\rm HXC}_l$ in OEP for a long chain with $m=20$ non-interacting
sites attached on each side of the five-site quantum dot with parameters $v_{\rm gate}/t = 0.5$ and $V/t = 0.5$.
In the figure, $v^{\rm HXC}_l$ is shown for the non-interacting sites $l = 1,\ldots,m$ and
for the whole chain in the inset.}
\label{voep}
\end{figure} 
\subsection{Results}
In the following we focus on the 
particle number in the interacting region, $N_{\rm dot}$,
as function of the gate voltage, $V_{\rm gate}$, comparing results for the three aforementioned
approaches LDA, OEP, and ED, respectively. The data are obtained 
for a chain of nine sites, i.e., the five-site quantum dot plus two non-interacting sites on each side of it.
Fig.~\ref{ndotV025} shows $N_{\rm dot}$ for weak interaction strength, $V/t = 0.25$.
The three curves nearly coincide with exception of the regions close to the steps, in particular
around $V_{\rm gate} = 0$. Here, as pointed out in Fig.~\ref{ndotV025a}, the step appears steeper in OEP and flatter in
LDA compared to ED. 

This trend continues at stronger interaction, $V/t = 0.5$, where in OEP the particle number
even jumps close to $V_{\rm gate} = 0$ with a small
hysteresis region of two stable solutions, while the LDA step flattens out even more,
as displayed in Fig.~\ref{ndotV05}.

In the strong coupling regime (Fig.~\ref{ndotV2}) where $V/t = 2$, comparison with the exact densities obtained from
DMRG calculations \cite{schmitteckert2008} shows that LDA and OEP fail completely, while
the ED results agree reasonably well with the exact data. Note however that the DMRG data are for a five-site quantum dot
with two non-interacting sites attached on the right and three on the left, and that the reservoirs are described
by a finite set of the order of 100 discrete levels instead of a continuum.
From the above observations we conclude that in DFT calculations for lattice models
LDA and OEP results are reliable in the weak interaction
regime. In particular, it can be shown that the OEP density profile is exact to 
linear order in the interaction strength $V$. 
On the other hand, for strongly correlated systems more sophisticated methods like 
the ED cluster approach are required, even for static properties.

As already mentioned in Sec.\ III.C, far from the interacting region the
potential $v_l^{\rm HXC}$ becomes very small. This is explicitly demonstrated in
Fig.~\ref{voep}: In the leads the Hartree-exchange-correlation potential,
$v_l^{\rm HXC}$, within OEP, is found to be about
three orders of magnitude smaller than in the interacting region (see inset).

\section{Transport}
DFT as presented in the previous sections is a ground state theory. However
generalizations are available which allow to
calculate densities at finite temperature and under non-equilibrium
conditions \cite{runge1984} and thus---via the
continuity equation---the current through
the quantum dot. 
The goal of this section is to calculate the DC-conductance of the
quantum dot when a small voltage difference is applied. 
We will extract the DC-conductance from a calculation of the dynamic
density response function, and the connection to the standard
Meir-Wingreen formula \cite{meir92} will be made.

\subsection{Conductance from density response}
We start with the current flowing from the left reservoir into the dot,
which is given by the time derivative of the particle number in
the reservoir,
\begin{equation}
I =  e  \dot N_L =  e  \sum_k \dot n_{kL}
,\end{equation}
i.e., $I(\omega) =  - i e \omega N_L(\omega)$, where $-e$ is the electron charge.
The frequency dependent variation in the particle number $N_L(\omega)$
appears as a response to a perturbation in the (single particle) Hamiltonian of the form
\begin{equation}
\delta \hat H^s = \sum_\alpha \hat n_\alpha \delta v_\alpha^s,
\end{equation}
where $\delta v_\alpha^s = \delta v_\alpha^{\rm ext} 
+ \delta v_\alpha^{\rm HXC} $ is the sum of an external potential and the induced
Hartree-exchange-correlation potential. The summation  $\alpha$ 
includes both reservoir, $\alpha = kL, kR$ and chain degrees of
freedom, $\alpha = l$. 
The variation of the density at site
$\beta$ is then
\begin{equation}
\delta n_\beta(\omega)  =  -i \sum_{\alpha}  \int \frac{d \epsilon}{2 \pi} 
{\cal G}_{\epsilon+ \omega}(\beta, \alpha) \delta v_\alpha^s(\omega)
{\cal G}_{\epsilon}(\alpha,\beta) 
,\end{equation}
where $\cal G_{\epsilon}(\alpha,\beta)$ is the (zero temperature)
Green's function of the single-particle Hamiltonian.
It is useful to distinguish the Green's function of the
reservoirs
from the Green's function of
the chain, and in the following we will use the symbols
$g_\epsilon(k L )$ for the (left) reservoir Green's function and
$G_\epsilon(l,l')$ for a chain Green's function. 
The latter is given by
\begin{equation}
G_\epsilon(l,l')  =   G^0_\epsilon(l,l') + \sum_{m} G^0_\epsilon(l,m) \Sigma_\epsilon(m) G_\epsilon(m,l')
,\end{equation}
were $G^0_\epsilon(l,l')$ is the bare Green's function, i.e., the one for $t_k= 0$,
and the self-energy $\Sigma_\epsilon(m)$ appears due to the coupling to the reservoirs.
For our model Hamiltonian $\Sigma_\epsilon(m)$ is
non-zero only on the first and last site of the chain, $m= 1, N$. The explicit expression 
for the first site is
\begin{equation}
\Sigma_\epsilon(1) = \sum_k | t_k |^2 g^0_{\epsilon}(kL)
,\end{equation}
where 
\begin{equation}
g^0_{\epsilon}(kL) = \frac{1}{\epsilon - \epsilon_k + i \delta\,{\rm sgn}(\epsilon - \mu )}, 
  \quad \delta = 0^+ 
\end{equation}
is the bare Green's function of state $k$ in the left lead.
For energies close to the chemical potential and for our special choice of
the couplings $t_k$ the self-energy then
assumes the value
$\Sigma_\epsilon(1) = - i t\,{\rm sgn}(\epsilon - \mu )$.
The Green's function for the states in the left reservoir finally is
\begin{equation}
g_{\epsilon }(kL) = g^0_{\epsilon }(kL) + g^0_{\epsilon } (kL) \,
t_k \,
G_\epsilon(1,1) \, t_k  \, g^0_{\epsilon }(kL)
.\end{equation}
The variation of the particle number in the reservoirs can be
represented graphically as
\begin{equation}
\includegraphics[width=0.95\linewidth]{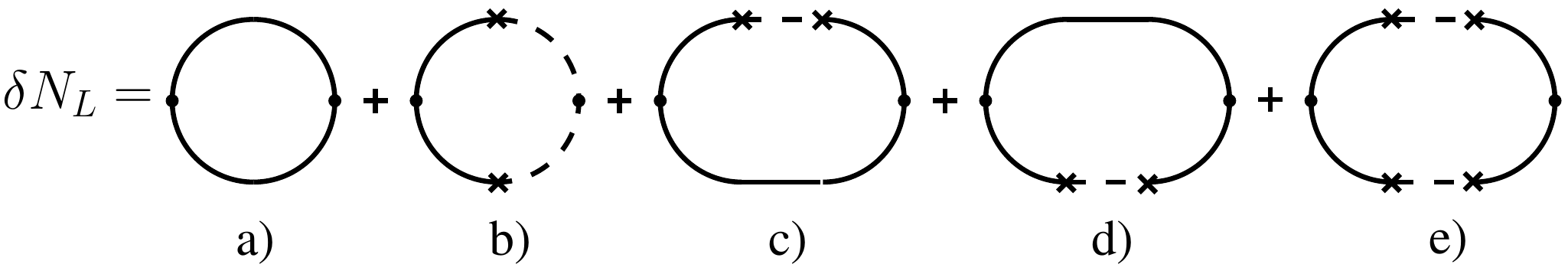} 
\end{equation}
where the full line denotes a reservoir Green's function,
$g^0_\epsilon (kL)$ or $g^0_\epsilon (kR)$, the broken line
is the chain Green's function $G_\epsilon(l,l')$ and the cross corresponds
to a hopping process between reservoir and chain.
In order to calculate the DC conductance of the system we have to
evaluate these diagrams for small but finite frequency, and we have to
identify contributions that diverge as $1/\omega$ as $\omega$ goes to
zero. Such divergences
are found in diagrams c), d) and e). In all three cases the diverging
contribution arises from the region in the $\epsilon$-integration where
$\epsilon <  \mu < \epsilon + \omega $. For diagram c), for instance,
the relevant contribution is
\begin{eqnarray} \label{eq30}
\delta N_L^{c)} & = & -i  \int_{\mu - \omega}^\mu \frac{d \epsilon}{2 \pi}
      G_{\epsilon+\omega}(1,1)  \\
&& \times \sum_k t_k^2
\left( \frac{1}{\epsilon + \omega - \epsilon_k + i \delta } \right)^2
       \frac{1}{\epsilon- \epsilon_k - i \delta} \delta v_{kL}^s \nonumber
.\end{eqnarray}
In the next step the $k$-summation in the second line is replaced by an
integral. Clearly the dominant contribution to the $k$-integration
comes from a small region around the Fermi energy. Assuming that
neither the
potential $\delta v_{kL}^s$ nor the coupling $t_k$ are singular around the Fermi momentum
we find $\sum_k (\dots ) = i \Gamma_L \delta v^s_{kL} /\omega^2$,
where the potential has to be evaluated at the Fermi energy.
The variation of the particle number in the left reservoir hence is
\begin{equation}
\delta N_L^{c)} = \frac{\delta v^s_{kL}}{2 \pi \omega} \Gamma_L G^R_\mu(1,1)
.\end{equation}
The retarded chain Green's function at the chemical potential,
$G^R_\mu(1,1)$, appears since
$\epsilon+ \omega > \mu$  in 
Eq.~(\ref{eq30}) and we consider the limit $\omega \to 0$.

Using similar arguments for the diagrams d) and e), we find
\begin{eqnarray}
\delta N_L^{d)} &= &-\frac{ \delta v_{kL}^s}{2 \pi \omega} \Gamma_L G^A_\mu(1,1)  \\
\delta N_L^{e)} &= & \frac{i \delta v_{kL}^s}{2 \pi \omega} \left[ \Gamma_L G^R_\mu(1,1)
\Gamma_L G^A_\mu(1,1) \right] \nonumber \\
&&  + \frac{i \delta v_{kR}^s}{2 \pi \omega} \left[ \Gamma_L G^R_\mu(1,N) \Gamma_R G^A_\mu(N,1) \right]
.\end{eqnarray}
Using finally the relation 
\begin{eqnarray}
G^R_\mu(1,1) -G^A_\mu(1,1)  &=  & -i G^R_\mu(1,1)\Gamma_L G^A_\mu(1,1)
\nonumber \\
            &&                 -i G^R_\mu(1,N)\Gamma_R G^A_\mu(N,1)
\end{eqnarray}
the complete singular contribution to the density in the reservoir
reads
\begin{equation}
\delta N_L = 
( \delta v_{kR}^s- \delta v_{kL}^s )
\frac{i}{2 \pi \omega}
 \Gamma_L G^R_\mu(1,N) \Gamma_R G^A_\mu(N,1) 
.\end{equation}
This enables us to write the current as the product of a conductance $G$ and a
voltage $U^{\rm tot}$,
\begin{equation}
I = -i e \omega \delta N_L = G U^{\rm tot} 
,\end{equation}
where the expression for the conductance agrees with the standard Meir-Wingreen formula
for non-interacting electrons,
\begin{equation} \label{eqConductance}
G = \frac{e^2}{2\pi \hbar} \Gamma_L G^R_\mu(1,L) \Gamma_R G^A_\mu(L,1) 
, \end{equation}
and the voltage is the sum of an externally applied voltage  and
an exchange-correlation contribution, $ U^{\rm tot} = U^{\rm ext} +
U^{\rm xc}$, with 
\begin{equation}
e U^{\rm ext} = \delta v^{\rm ext}_{kR} - \delta v^{\rm ext}_{kL},
\quad
e U^{\rm xc} = \delta v^{\rm HXC}_{kR} - \delta v^{\rm HXC}_{kL}
.\end{equation}

Notice that only the exchange-correlation potentials in the reservoirs
but not in the chain contribute to the total voltage.
Our result is thus consistent with  [\onlinecite{stefanucci2004}] where it has
been shown that the exact current can be expressed in terms of  a
Landauer-type formula \cite{landauer1957} in which the electro-chemical potential of the
leads is shifted by the voltage-induced variation of the exchange
correlation potential, and with the statements of Refs.\ 
[\onlinecite{mera2010,koentopp2006}] that static
DFT gives the exact linear-response conductance provided that the
dynamic exchange-correlation potential vanishes deep inside the leads.

Notice that $U^{\rm xc}$ is a
purely dynamic effect that cannot be captured by any adiabatic
approximation.
However, $U^{\rm xc}$ can be assessed via reverse engineering: If we know 
the exact static densities and the exact conductance, then the
ratio of the exact conductance and the DFT conductance, Eq.~(\ref{eqConductance}), 
is equal to the ratio of the total and the external voltage, 
$G^{\rm exact}/G = U^{\rm tot}/U^{\rm ext}$.

\subsection{Results}
\begin{figure}
\includegraphics[width=8.5cm]{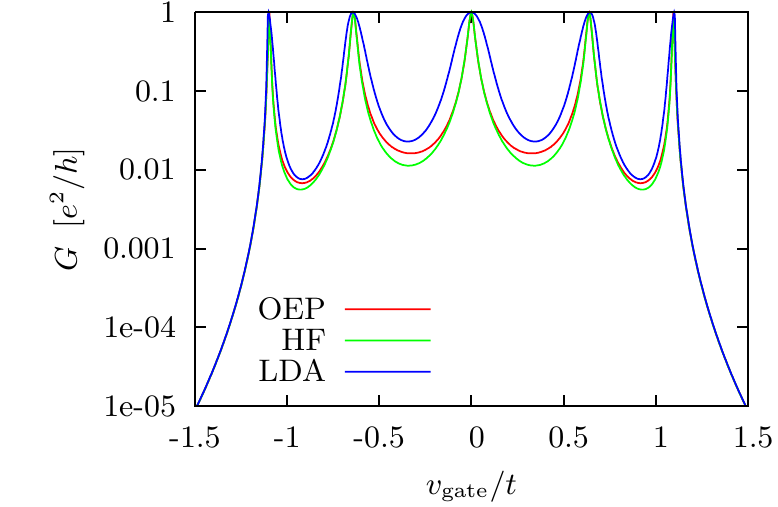} 
\caption{Conductance $G$ as function of $v_{\rm gate}$ for $V/t =
0.25$. The OEP and LDA curves are obtained within static density
functional theory as explained in the text. ``HF'' corresponds to a
self-consistent Hartree-Fock calculation.}
\label{gV025}
\end{figure}

Figure \ref{gV025} shows the conductance as a function of the gate
voltage in the case of a weak interaction $V=0.25t$. One observes
five resonances at the gate voltages where the particle number on the
dot changes, compare Fig.~\ref{ndotV025}. LDA overestimates the width of
the resonances just as it overestimates the width of the steps in the
particle number. We also include OEP and Hartree-Fock (HF) results in the
figure. For weak interaction both methods predict identical charge
densities, for the present interaction strength the difference in particle number on
the dot is for the two methods less than $7\cdot10^{-3}$. Also the conductances
are close to each other. 

\begin{figure}
\includegraphics[width=8.5cm]{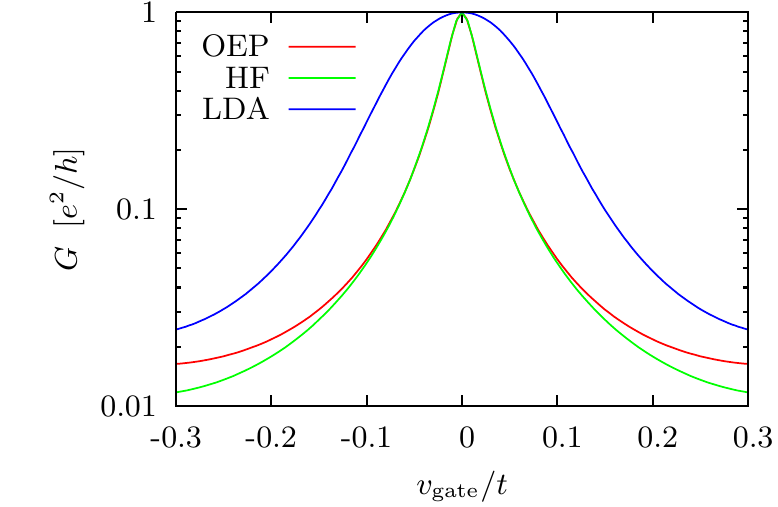} 
\caption{Conductance $G$ as function of $v_{\rm gate}$ close to the central peak for $V/t = 0.25$.}
\label{gV025a}
\end{figure}

In Fig.~\ref{gV025a} we show the region close
to zero gate voltage in more detail. Near the resonance HF and OEP are
almost indistinguishable, however far from the resonance a
significantly different conductance is found.
This difference is due to the exchange-correlation
contribution to the voltage, $U^{\rm xc}$.
To substantiate this point we analyze the correction to the
conductance to first order in the interaction strength $V$. 
In this case HF yields the exact conductance. 

\begin{figure}
\includegraphics[width=8.5cm]{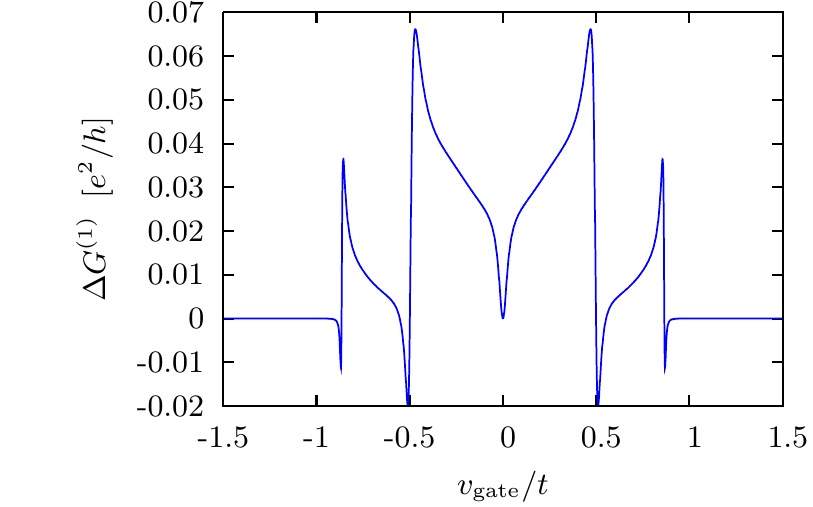}
\caption{Linear contribution $G^{(1)}$ of the expansion of the conductance
$G = G^{(0)} + G^{(1)} V/t + G^{(2)} (V/t)^2 + \ldots $ in powers of the interaction
strength. The difference between the OEP result and the exact HF conductance,
$\Delta G^{(1)} = G^{(1)}_{\rm OEP} - G^{(1)}_{\rm HF}$, is
plotted as function of $v_{\rm gate}$.}
\label{deltag}
\end{figure}

However, as
demonstrated in Fig.~\ref{deltag} the conductances obtained by OEP and
HF differ---although the densities are identical. Figure \ref{deltag} has been obtained
for a chain length $N= 9$, but we have checked that even for much
longer chains (up to $N=25$) there is no visible change in the results.
This means that $U^{\rm xc}$ remains nonzero even far from the
interacting region.

\begin{figure}
\includegraphics[width=8.5cm]{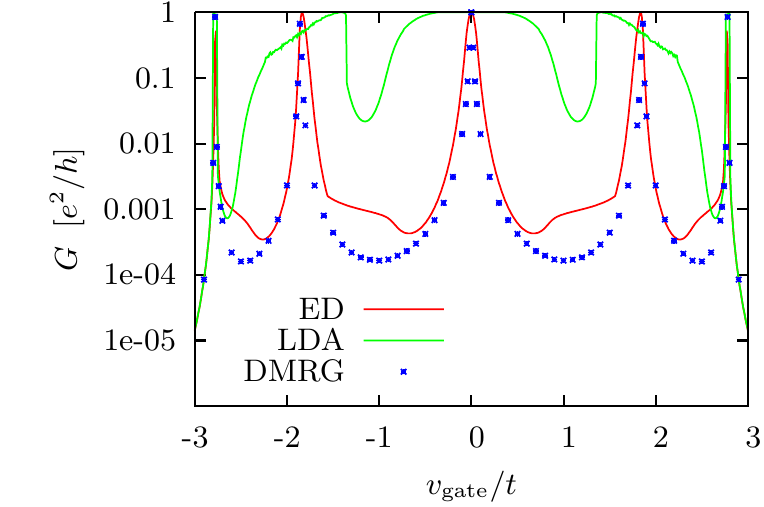} 
\caption{Conductance $G$ as function of $v_{\rm gate}$ for $V/t = 2$;
the DMRG data are taken from Ref.\  [\onlinecite{schmitteckert2008}].}
\label{gV2}
\end{figure}

Figure~\ref{gV2} shows the conductance for the relatively strong
interaction strength $V=2t$, comparing LDA, ED and the numerically exact
conductance obtained with the DMRG in Ref.~[\onlinecite{schmitteckert2008}].
Although the position of the resonances is not bad in LDA the method
predicts conductances that differ several orders of magnitude from the
exact results. The ED results are considerably better. Near the 
resonances ED predicts conductances that are close to the DMRG
values.

\section{Summary and Conclusions}
We studied the ground state density profile and the conductance of a
model quantum dot comparing DFT and exact results. 
The electron density in the ground state can be obtained reliably using non-local 
exchange-correlation potentials. While for weak interaction the OEP approach gives 
good results, in the case of intermediate or
strong interaction strengths a non-local potential extracted from the exact 
diagonalization of small clusters (DFT+ED) works well.

For the conductance our results are not so clear-cut. In our simple model we 
find five well separated resonances as a function of the gate voltage. Static DFT 
reproduces very well the position and the width of these resonances. 
However, in contrast to model cases where the Friedel sum rule allows to express 
the conductance in terms of the equilibrium charge density, compare
Ref.~[\onlinecite{mera2010}], in our model system such a sum rule is not valid.
As a consequence the DFT conductance is {\em not} exact, and we find pronounced 
deviations in the conductance valleys between the resonances. 

As the origin of these discrepancies we have identified a dynamical 
exchange-correlation correction to the applied voltage, which is non-zero even
when the reservoirs are far from the interacting region.
We believe that this finding is related to ``ultra-non-locality'' which
is an inherent problem in time-dependent density functional theory 
\cite{vignale1996,sai2005}. 

We thank C.~Schuster for fruitful discussions as well as 
the Deutsche Forschungsgemeinschaft (TRR80) for financial support.

\end{document}